\documentclass[11pt,a4paper]{article}
\usepackage{jheppub}

\usepackage{amsmath,amssymb}
\usepackage{tikz,tikz-feynman}
\usepackage{graphicx,float,subfigure}
\usepackage{ulem}

\usepackage{hyperref}

\begin{document}

\title{Uniqueness of gravitational constant at low energies from the connection between spin-2 and spin-0 sectors}

\author{Duojie Jimu and}
\author{Tomislav Prokopec}

\affiliation{Institute for Theoretical Physics, Spinoza Institute \& EMME$\Phi$, Utrecht University, Princetonplein 5, 3584 CC Utrecht, The Netherlands}

\emailAdd{jm.dj@outlook.com}
\emailAdd{T.Prokopec@uu.nl}

\abstract{
The fact that graviton propagator contains not only one but two tensorial components excludes a unique definition of the running behavior of the gravitational constant, while at low energies gravitation is characterized solely by Newton's constant. How these two facts are reconciled when massive quantum fields are present remains unanswered. In this work, by non-minimally coupling gravity to a one-loop massive scalar, we show that this potential conflict is resolved by the non-trivial equivalence between the residues of the two propagator components. Such equivalence, crucial for the validity of the Appelquist-Carazzone decoupling theorem, is based on a rather subtle connection between the spin-2 and spin-0 sectors of the propagator. It is verified that this connection also makes the two quantum-corrected gravitational potentials be characterized by the same gravitational constant at large distances. In addition, we find that the potentials in our case as well as the quantum-corrected Coulomb potential can be expressed concisely in a unified formulation. By comparing these results with experiments, we establish a new upper bound on the magnitude of the non-minimal coupling parameter $\xi$.
}

\maketitle
%
\section{Introduction}
\label{Introduction}
Unlike the coupling constants in the Standard Model, there is not a unique way to define the running behavior of the gravitational constant $G$ within the framework of perturbative quantum gravity (pQG), as pointed out in~\cite{Anber:2011ut}. We need different definitions to accommodate the energy dependence of different gravitational scattering amplitudes, and this may reflect our limited access to the full quantum nature of gravity. In fact, we can already see such issue arising from the structure of the graviton propagator.

In general, the graviton propagator contains two orthogonal tensorial components referred to as the spin-2 and spin-0 components respectively. As a result, given a dressed graviton propagator which encodes certain loop effects, we may define the running behavior of $G$ according to either component, and this usually leads to different behavior. On the other hand, classical gravity says there should be a unique gravitational constant, namely Newton's constant, at low energies. To reconcile these two facts, we need to make sure the different definitions of the running behavior all run to the same value at low energies. In~\cite{Anber:2011ut}, the authors argued that this is exactly the case in pure gravity on the ground of its one-loop finiteness.

However, the situation dramatically changes when we include the loop effects of massive fields. This is because the counterterms in the previous case is now joined by a new term $m^{2}R$ where $m$ is the mass of the particle running in the loop and $R$ is the Ricci scalar, and this term does affect low energy physics. In this paper, we show that, even in this case, the definition of the low energy gravitational constant is unique, but for a more delicate reason based on the hidden connection between the spin-2 and spin-0 sectors.

The system we consider is gravity non-minimally coupled to a one-loop massive scalar. We will demonstrate the uniqueness at two levels. The first is field theoretic and concerns the structure of the dressed graviton propagator. We show that the two tensorial components of it have non-trivial yet {\it equal} residues at the massless pole. This amounts to say the two different running gravitational couplings run to the same value at low energies. The second level is more related to the physical quantities we can actually measure, where we compute the two quantum-corrected gravitational potentials and show that, in their asymptotic values at large distances: $\Phi(r)\sim-\frac{GM}{r}(1+c_{\Phi})$ and $\Psi(r)\sim-\frac{GM}{r}(1+c_{\Psi})$, the two constants $c_{\Phi}$ and $c_{\Psi}$ are {\it equal}. This shows in an operational way the uniqueness of the low energy gravitational constant.

These results suggest the gravity-massive scalar system as an illuminating example for the Appelquist-Carazzone decoupling theorem~\cite{Appelquist:1974tg}~\footnote{We thank the anonymous referee for pointing out the connection of our work with this theorem.}, as the massive scalar not only decouples individually from every degrees of freedom of gravity in the infrared, but also guarantees that the pieces of its leftovers could fit together to restore classical gravity. This provides a novel check on the self-consistency of pQG and gives more credence to it as the low energy description of quantum gravity,
and moreover it renders support to Donoghue's conjecture~\cite{Donoghue:2012zc} that gravity
in the infrared (on scales far below the Planck scale) can be viewed as an effective field theory.

In the meantime, we fill in a gap in the literature~\cite{Burns:2014bva, Frob:2016xte} regarding the analytic formula of the quantum-corrected gravitational potentials in our case. We apply the techniques from quantum electrodynamics (QED) to obtain the formula and this in turn reveals the common structures shared by the force laws in pQG and QED. Finally, by comparing the non-zero gravitational slip deduced from the potentials with the experiments, we establish a new constraint on the magnitude of the non-minimal coupling parameter $\xi$.

The structure of the paper is as follows. In Section~\ref{Preparation: deriving the dressed graviton propagator}, we derive the dressed graviton propagator encoding the loop effects of the massive scalar. This section is prepared for Section~\ref{The spin-2 and spin-0 residues match up} where we show the equivalence of the two residues. Next, in Section~\ref{Preparation: deriving the quantum-corrected gravitational potentials}, we calculate the quantum corrections to the two gravitational potentials, which is prepared for Section~\ref{Corrections to the two potentials converge at large distances} to show their convergence at large distances. Section~\ref{Properties of the gravitational potentials in position space} is devoted to a detailed study of the quantum-corrected gravitational potentials about their theoretical properties and experimental consequences. We summarize our findings in Section~\ref{Discussion}.

We use $(-,+,+,+)$ for the metric signature. The Riemann tensor is defined as
$R^{\rho}_{\ \ \sigma\mu\nu}=\partial_{\mu}\Gamma^{\rho}_{\nu\sigma}+\Gamma^\rho_{\;\mu\lambda}\Gamma^\lambda_{\;\nu\sigma}
-(\mu\leftrightarrow \nu)$. The graviton field is defined as the perturbation to the inverse metric,
{\it i.e.} $g^{\mu\nu}(x)=\eta^{\mu\nu}+\kappa h^{\mu\nu}(x)$, with $\kappa=\sqrt{16\pi G}$ and $\eta_{\mu\nu}$ being the Minkowski metric. We use natural units $\hbar=1=c$.
\section{Preparation: deriving the dressed graviton propagator}
\label{Preparation: deriving the dressed graviton propagator}
We start by setting up the framework for perturbative calculation. Our action contains the Einstein-Hilbert term plus a non-minimally coupled real massive scalar,
\begin{equation}\label{original_action}
S=\frac{1}{16\pi G}\int \mathrm{d}^{D}x\sqrt{-g}R+\int \mathrm{d}^{D}x\sqrt{-g}\left(-\frac{1}{2}g^{\mu\nu}\partial_{\mu}\phi\partial_{\nu}\phi-\frac{1}{2}m^{2}\phi^{2}-\frac{1}{2}\xi R\phi^{2}     \right).
\end{equation}
Similar to \cite{Capper:1973bk}, our graviton field $h^{\mu\nu}$ is defined as the perturbation to the inverse metric,
\begin{equation}
g^{\mu\nu}=\eta^{\mu\nu}+\kappa h^{\mu\nu},
\end{equation}
where $\eta^{\mu\nu}$ is the Minkowski metric and $\kappa \equiv \sqrt{16\pi G}$. Indices are raised or lowered by $\eta_{\mu\nu}$, and the trace of $h_{\mu\nu}$ is denoted by $h \equiv h_{\mu\nu}\eta^{\mu\nu}$. To study the one-loop effects induced by the scalar, we only need to expand the action to quadratic order in $h_{\mu\nu}$. The result reads:
\begin{equation}\label{expanded_action}
\begin{aligned}
S&=S_{h}+S_{\phi}+S_{3}+S_{4},\\
S_{h}&=\int \mathrm{d}^{D}x \left[-\frac{1}{4}\partial_{\alpha}h_{\mu\nu}\partial^{\alpha}h^{\mu\nu}+\frac{1}{4}\partial_{\mu}h\partial^{\mu}h+\frac{1}{2}\partial_{\mu}h^{\mu\nu}(\partial^{\alpha}h_{\nu\alpha}-\partial_{\nu}h) \right],\\
S_{\phi}&=\int \mathrm{d}^{D}x \left(-\frac{1}{2}\partial_{\mu}\phi\partial^{\mu}\phi-\frac{1}{2}m^{2}\phi^{2} \right),\\
S_{3}&=-\frac{\kappa}{2}\int \mathrm{d}^{D}x\,T^{(\phi)}_{\mu\nu}h^{\mu\nu},\\
S_{4}&=\frac{\kappa^{2}}{4}\int \mathrm{d}^{D}x \left[ \partial_{\mu}\phi\partial_{\nu}\phi h^{\mu\nu}h-\left(\frac{h^{2}}{4}+\frac{h_{\mu\nu}h^{\mu\nu}}{2}\right)(\partial_{\mu}\phi\partial^{\mu}\phi+m^{2}\phi^2)+\xi\phi^{2}h^{\mu\nu}\mathcal{L}_{\mu\nu\rho\sigma}h^{\rho\sigma} \right],
\end{aligned}
\end{equation}
in which $T^{(\phi)}_{\mu\nu}$ denotes the energy-momentum tensor of the scalar field on Minkowski background,
\begin{equation}\label{Tuv}
T^{(\phi)}_{\mu\nu}=\partial_{\mu}\phi\partial_{\nu}\phi-\frac{\eta_{\mu\nu}}{2}(\partial_{\alpha}\phi\partial^{\alpha}\phi+m^{2}\phi^{2})+\xi\Big[\eta_{\mu\nu}\partial^{2}(\phi^{2})-\partial_{\mu}\partial_{\nu}(\phi^{2})\Big],
\end{equation}
and $\mathcal{L}_{\mu\nu\rho\sigma}$ is the flat space Lichnerowicz operator,
\begin{equation}\label{lich}
\mathcal{L}_{\mu \nu \rho \sigma}=\frac{1}{2}(\eta_{\mu \nu}\eta_{\rho \sigma}\partial^{2}-\eta_{\rho \sigma}\partial_{\mu}\partial_{\nu}-\eta_{\mu \nu}\partial_{\rho}\partial_{\sigma}-\eta_{\mu (\rho}\eta_{\sigma) \nu}\partial^{2}+2\partial_{(\mu}\eta_{\nu) (\rho}\partial_{\sigma)}).
\end{equation}
Here $S_{h}$ is the quadratic action of graviton, $S_{\phi}$ is the action of a free scalar, $S_{3}$ and $S_{4}$ 
are the three-point and four-point interaction terms, where in $S_{4}$ we have dropped a term in the form $\xi \partial(\phi^{2})h\partial h$, since its contribution to the one-loop Feynman diagrams vanishes.

The dressed graviton propagator $i\big[{}_{\mu\nu}\Delta_{\rho\sigma}^{(1)}\big]$ which encodes the one-loop effects of the scalar can be expressed schematically as the sum in Figure~\ref{figure_pro}.
\begin{figure}[H]
\begin{equation*}
i\big[{}_{\mu\nu}\Delta_{\rho\sigma}^{(1)}\big]=
\begin{tikzpicture}
\begin{feynman}
\vertex (a);
\vertex [right=of a] (b);
\diagram{
(a) -- [gluon] (b)
};
\end{feynman}
\end{tikzpicture}
\ +\ 
\begin{tikzpicture}[baseline=-0.6ex]
\begin{feynman}
\vertex (a) at (0,0);
\vertex[blob] (b) at (1.2, 0) {$\ \ \ \ $};
\vertex (c) at (2.4,0);
\diagram* {
(a) -- [gluon] (b) -- [gluon] (c);
};
\end{feynman}
\end{tikzpicture}
\ +\ 
\begin{tikzpicture}[baseline=-0.6ex]
\begin{feynman}
\vertex (a) at (-0.2,0);
\vertex[blob] (b) at (1, 0) {$\ \ \ \ $};
\vertex[blob] (c) at (2.6, 0) {$\ \ \ \ $};
\vertex (d) at (3.8,0);
\diagram* {
(a) -- [gluon] (b) -- [gluon] (c) -- [gluon] (d);
};
\end{feynman}
\end{tikzpicture}
\ +\ \cdots
\end{equation*}
\caption{The dressed propagator given by a summation over the bubble diagrams with increasing number of self-energy insertions.
\label{figure_pro}}
\end{figure}
\noindent Here the bubble and the coiled line represent the scalar induced graviton self-energy and the bare graviton propagator respectively. We first compute the self-energy.
\subsection{Manifestly transverse self-energy}
\label{Manifestly transverse self-energy}
The graviton self-energy contains the three diagrams in Figure~\ref{figure_se}.
\begin{figure}[H]
\centering
\subfigure[\,Non-local diagram]{
\begin{tikzpicture}
\begin{feynman}
\vertex (a);
\vertex [left=1.5 cm of a] (e);
\vertex [above right=1cm of a] (b);
\vertex [below right=1cm of b] (c);
\vertex [below left=1cm of c] (d);
\vertex [right=1.5cm of c] (f);
\diagram{
(e) -- [gluon, momentum' = $k$, edge label=$\mu\nu$] (a);
(a) -- [quarter left] (b) -- [quarter left] (c) -- [quarter left] (d) -- [quarter left] (a);
(c) -- [gluon, momentum' = $k$, edge label=$\rho\sigma$] (f);
};
\end{feynman}
\end{tikzpicture}
}
\qquad
\subfigure[\,Local diagram]{
\begin{tikzpicture}
\begin{feynman}
\vertex (a);
\vertex [left=2cm of a] (e);
\vertex [right=2cm of a] (f);
\vertex [above left=1cm of a] (b);
\vertex [above right=1cm of b] (c);
\vertex [below right=1cm of c] (d);
\diagram{
(e) -- [gluon, momentum' = $k$, edge label=$\mu\nu$] (a);
(a) -- [quarter left] (b) -- [quarter left] (c) -- [quarter left] (d) -- [quarter left] (a);
(a) -- [gluon, momentum' = $k$, edge label=$\rho\sigma$] (f);
};
\end{feynman}
\end{tikzpicture}
}
\qquad
\subfigure[\,Counterterms]{
\begin{tikzpicture}
\begin{feynman}
\vertex (a);
\vertex [left=of a] (b);
\vertex [right=of a] (c);
\diagram{
(b) -- [gluon, momentum'=$k$, edge label=$\mu\nu$] a [crossed dot] -- [gluon, momentum'=$k$, edge label=$\rho\sigma$] (c)
};
\end{feynman}
\end{tikzpicture}
}
\hspace{0.5cm}
\caption{Feynman diagrams contributing to the graviton self-energy. Coiled lines and solid lines represent graviton and the scalar, respectively.
\label{figure_se}}
\end{figure}
\noindent By using the Feynman rules given in Appendix~\ref{Feynman rules and amplitudes}, one can evaluate the amplitudes of diagrams (a) and (b) and add them together to reach the primitive one-loop self-energy,
\begin{equation}
\begin{aligned}
-i\big[{}_{\mu\nu}\Sigma_{\rho\sigma}\big]&=\kappa^{2}\left[\frac{4(D-2)m^{2}-(D^{2}-2D-2)k^2}{16(D^{2}-1)}P_{\mu\nu}P_{\rho\sigma}+\frac{4(D-2)m^{2}-k^{2}}{8(D^{2}-1)}P_{\mu\nu\rho\sigma}\right]I_{1}\\
&\quad \ +\kappa^{2}\left[\frac{(Dk^{2}-4m^{2})^{2}-2(D+1)k^4}{32(D^{2}-1)}P_{\mu\nu}P_{\rho\sigma}+\frac{(k^{2}+4m^{2})^{2}}{16(D^2-1)}P_{\mu\nu\rho\sigma}\right]I_{2}\\
&\quad \ +\kappa^{2}\xi\left[ \frac{(D-3)k^{2}}{4(D-1)}P_{\mu\nu}P_{\rho\sigma}+\frac{k^{2}}{4}P_{\mu\nu\rho\sigma} \right]I_{1}\\
&\quad \ +\kappa^{2}\xi\left[ \frac{\xi k^{4}}{2}-\frac{(D-2)k^{4}}{4(D-1)}+\frac{m^{2}k^{2}}{D-1}  \right]P_{\mu\nu}P_{\rho\sigma}I_{2}\\
&\quad \ -\frac{\kappa^{2}m^{2}}{4D}(\eta_{\mu\nu}\eta_{\rho\sigma}+2\eta_{\mu(\rho}\eta_{\sigma)\nu})I_{1}.
\end{aligned}
\label{self energy: primitive}
\end{equation}
where the transverse projectors $P_{\mu\nu}$ and $P_{\mu\nu\rho\sigma}$ are defined as,
\begin{equation}\label{projector_def}
P_{\mu\nu} \equiv \eta_{\mu\nu}-\frac{k_{\mu}k_{\nu}}{k^{2}}\,,\qquad \qquad P_{\mu\nu\rho\sigma} \equiv P_{\mu(\rho}P_{\sigma)\nu}\,,
\end{equation}
and the integrals $I_{1}$ and $I_{2}$ are defined as,
\begin{equation}\label{I1I2_def}
\begin{aligned}
I_{1}& \equiv \int \frac{\mathrm{d}^{D}p}{(2\pi)^{D}} \frac{1}{p^{2}+m^{2}}\,,\\
\qquad \qquad I_{2}& \equiv \int \frac{\mathrm{d}^{D}p}{(2\pi)^{D}} \frac{1}{(p^{2}+m^{2})[(p+k)^{2}+m^{2}]}\,.
\end{aligned}
\end{equation}
We see that, unlike the vacuum polarization in QED, the primitive graviton self-energy is not transverse due to the last term in~\eqref{self energy: primitive}. This can be traced back to the energy-momentum tensor of the scalar which can curve the space and spoil our flat space assumption, unless we introduce a specific cosmological constant to compensate for it. When the contribution from this cosmological constant is included, the primitive self-energy will become transverse, as we will see later.

The divergences in the primitive self-energy~(\ref{self energy: primitive}) can be removed by the following counterterms~\cite{Geist:1973my,Gorbar:2002pw,Kavanagh:2023},
\begin{equation}\label{counterterms}
\Delta S=\int d^{D}x\sqrt{-g}\left( c_{1}C_{\mu \nu \rho \sigma}C^{\mu \nu \rho \sigma}+c_{2}R^{2}+c_{3}R+\Lambda \right),
\end{equation}
with
\begin{equation}\label{c_123}
\begin{aligned}
c_{1}&=\frac{\Omega\mu^{D-4}}{240(D\!-\!4)}+c_{1\rm f},\\
c_{2}&=\frac{\Omega\mu^{D-4}}{4(D\!-\!4)}\left(\xi-\frac{1}{6}\right)^{2}+c_{2\rm f},\\
c_{3}&=\frac{\Omega m^{2}\mu^{D-4}}{2(D\!-\!4)}\left(\xi-\frac{1}{6}\right)+c_{3\rm f},\\
\Lambda&=\frac{\Omega m^{4}}{4}\left( \frac{\mu^{D-4}}{D\!-\!4}+\Gamma_{E}-\frac{1}{4} \right).
\end{aligned}
\end{equation}
Here $C_{\mu\nu\rho\sigma}$ is the Weyl tensor, $\mu$ is an arbitrary energy scale introduced for dimensional regularization, 
and we used shorthand notations,
\begin{equation}\label{shorthand_def}
\begin{aligned}
\qquad
\Omega \equiv \frac{1}{8\pi^{2}},\qquad \qquad
\Gamma_{E} \equiv \frac{1}{2}\left[\ln\left(\frac{m^{2}}{4\pi \mu^{2}}\right)+\gamma_{E}-1\right].
\end{aligned}
\end{equation}
In~\eqref{c_123}, $c_{1\rm f}$, $c_{2\rm f}$ and $c_{3\rm f}$ are finite counterterms and their values depend on the choice of renormalization scheme. The cosmological constant counterterm $\Lambda$ is somewhat special as it is completely fixed. As mentioned above, in addition to removing the divergences, $\Lambda$ also does the following jobs~\cite{Capper:1973bk,Burns:2014bva}:
\begin{enumerate}
\item[(1)]Its contribution to the expectation value of the total energy-momentum tensor cancels the one-loop contribution of the scalar field so that spacetime remains flat;
\item[(2)]As another manifestation of (1), its one-point diagram cancels the tadpole diagram and ensures $\langle\hat h_{\mu\nu}(x) \rangle=0$; 
\item[(3)]Its contribution to the graviton self-energy is essential for the latter to be transverse.
\end{enumerate}

By using the Feynman rules given in Appendix~\ref{Feynman rules and amplitudes}, 
we can evaluate the amplitude of the counterterm diagram in Figure~\ref{figure_se}, combine it with the primitive self-energy \eqref{self energy: primitive}, and reach the manifestly transverse, renormalized graviton self-energy,
\begin{equation}\label{finite_se}
-i\big[{}_{\mu\nu}\Sigma_{\rho\sigma}\big]=i\kappa^{2}AP_{\mu\nu}P_{\rho\sigma}+i\kappa^{2}BQ_{\mu\nu\rho\sigma}\,,
\end{equation}
with the transverse projector $Q_{\mu\nu\rho\sigma}$ defined as, 
\begin{equation}
Q_{\mu\nu\rho\sigma} \equiv 
 P_{\mu(\rho}P_{\sigma)\nu}-\frac{P_{\mu\nu}P_{\rho\sigma}}{D-1}
  \;\;\stackrel{D\rightarrow 4}{\longrightarrow}\;\;
   P_{\mu(\rho}P_{\sigma)\nu}-\frac{P_{\mu\nu}P_{\rho\sigma}}{3}
  \,,
\end{equation}
and the form factors given by:
\begin{equation}\label{AB}
\begin{aligned}
A&=-\frac{\Omega L(k^{2})}{4}\left[ k^{2}\left(\xi-\frac{1}{6}\right)+\frac{m^{2}}{3} \right]^{2}\\
&\quad \ +k^{4}\left[ -\frac{\Omega(\Gamma_{E}+\frac{1}{2})}{2}\left(\xi-\frac{1}{6}\right)^{2}+\frac{\Omega}{36}\left( \xi-\frac{1}{6} \right)+2c_{2\rm f} \right]\\
&\quad \ +k^{2}\left[ -\frac{\Omega m^{2}\Gamma_{E}}{6}\left( \xi-\frac{1}{6} \right)+\frac{\Omega m^{2}}{216}+\frac{c_{3\rm f}}{3} \right],\\
\ \\
B&=-\frac{\Omega L(k^{2})}{480}(k^{2}+4m^{2})^{2}+k^{4}\left( -\frac{\Omega \Gamma_{E}}{240}+\frac{\Omega}{450}+c_{1\rm f} \right)\\
&\quad \ +k^{2}\left[ \frac{\Omega m^{2}\Gamma_{E}}{4}\left( \xi-\frac{1}{6} \right)+\frac{\Omega m^{2}}{180}-\frac{c_{3\rm f}}{2} \right],
\end{aligned}
\end{equation}
in which $L(k^{2})$, encoding the non-local effects, is defined as,
\begin{equation}\label{L_def}
L(k^{2})=\int_{0}^{1}\mathrm{d}\alpha \ln\left[1+\frac{k^{2}}{m^{2}}\alpha(1-\alpha)\right]=-2+\sqrt{1+\frac{4m^2}{k^2}}\ln\left[ \frac{\sqrt{1+\frac{4m^2}{k^2}}+1}{\sqrt{1+\frac{4m^2}{k^2}}-1} \right].
\end{equation}
In \eqref{L_def}, the logarithm and square root should be understood as complex functions whose imaginary parts are uniquely fixed by the $i\epsilon$ prescription which can be restored by the substitution $k^{2} \to k^{2}-i\epsilon$.

$P_{\mu\nu}P_{\rho\sigma}$ and $Q_{\mu\nu\rho\sigma}$ represent the spin-0 and spin-2 degrees of freedom of the self-energy respectively. They satisfy the following identities:
\begin{equation}\label{PQ_identities}
\begin{aligned}
P_{\mu\alpha}P^{\alpha}_{\ \ \nu}&=P_{\mu\nu}, \qquad \qquad Q_{\mu\nu\alpha\beta}Q^{\alpha\beta}_{\ \ \ \rho\sigma}=Q_{\mu\nu\rho\sigma}, \qquad \qquad
Q_{\mu\nu\rho\sigma}P^{\rho\sigma}=0,\\
P_{\mu\nu}\eta^{\mu\nu}&=D-1, \qquad \qquad \ \, Q_{\mu\nu\rho\sigma}\eta^{\rho\sigma}=0.
\end{aligned}
\end{equation}
The property that $Q_{\mu\nu\rho\sigma}$ is orthogonal to $P_{\mu\nu}$ will significantly simplify the contraction of indices in the next subsection.
\subsection{Resumming diagrams}
\label{Resumming diagrams}
Now that we have obtained the self-energy, the other ingredient we need to perform the summation in Figure~\ref{figure_pro} is the bare graviton propagator. We shall work in the general covariant gauge which includes the usual harmonic or Landau gauge as special cases and is obtained by adding the following gauge fixing term to the original action $S_{h}$ in~\eqref{expanded_action},
\begin{equation}
S_{\rm GF}[h_{\mu\nu}] = \int {\rm d}^D x \,{\cal L}_{\rm GF}
\,,\qquad
{\cal L}_{\rm GF}=-\frac{1}{2\alpha}\Big(\partial_{\mu}h^{\mu}_{\;\gamma}-\frac{\beta}{2}\partial_\gamma h\Big)\eta^{\gamma\delta}
    \Big(\partial_{\nu}h^{\nu}_{\;\delta}-\frac{\beta}{2}\partial_\delta h\Big),
\label{gauge fixing action}
\end{equation}
where $\alpha,\beta\in \mathbb{R}$ are real parameters. The resulting bare graviton propagator $i\big[{}_{\mu\nu}\Delta_{\rho\sigma}^{(0)}\big]$ is derived as usual by calculating the inverse of the quadratic action and the result reads (in $D$ dimensions)~\cite{Capper:1979ej,Miao:2017feh},
\begin{equation}\label{graviton propagator: ab gauge}
\begin{aligned}
i\big[{}_{\mu\nu}\Delta_{\rho\sigma}^{(0)}\big] &=\frac{2i}{k^{2}} \Bigg[
\frac{1}{(D-1)(D-2)}\left(P_{\mu\nu} - \frac{(D\!-\!1)\beta}{\beta-2}\frac{k_\mu k_\nu}{k^2}\right)
           \left(P_{\rho\sigma} - \frac{(D\!-\!1)\beta}{\beta-2}\frac{k_\rho k_\sigma}{k^2}\right)\\
          &\qquad \quad \ 
          -Q_{\mu\nu\rho\sigma}-2\alpha \frac{k_{(\mu} P_{\nu)(\rho}k_{\sigma)}}{k^2}
           -\frac{2\alpha}{(\beta-2)^2} \frac{k_{\mu} k_{\nu}k_{\rho}k_{\sigma}}{k^4}
  \Bigg].
\end{aligned}
\end{equation}

Now we have all the ingredients we need to perform the summation in Figure~\ref{figure_pro}. Using the identities in~\eqref{PQ_identities} and the transversality of $P_{\mu\nu}$ and $Q_{\mu\nu\rho\sigma}$ repeatedly, we find that the summation corresponds to a geometric series as usual and can be straightforwardly performed to give the one-loop dressed graviton propagator,
\begin{equation}\label{dressed graviton propagator}
\begin{aligned}
i\big[{}_{\mu\nu}&\Delta_{\rho\sigma}^{(1)}\big]\\
&=i\big[{}_{\mu\nu}\Delta_{\rho\sigma}^{(0)}\big]+i\big[{}_{\mu\nu}\Delta_{\alpha\beta}^{(0)}\big]\big(-i\big[{}^{\alpha\beta}\Sigma^{\gamma\lambda}\big]\big)i\big[{}_{\gamma\lambda}\Delta_{\rho\sigma}^{(0)}\big]+\cdots\\
&=\frac{2i}{k^{2}}\Bigg[
 \frac{1}{(D\!-\!1)(D\!-\!2)}\frac{1}{1+\frac{2(D-1)\kappa^{2}}{(D-2)k^{2}}A}\left(P_{\mu\nu} - \frac{(D\!-\!1)\beta}{\beta-2}\frac{k_\mu k_\nu}{k^2}\right)
           \left(P_{\rho\sigma} - \frac{(D\!-\!1)\beta}{\beta-2}\frac{k_\rho k_\sigma}{k^2}\right)\\
           &\qquad \quad \ 
           -\frac{1}{1-\frac{2\kappa^{2}}{k^{2}}B}Q_{\mu\nu\rho\sigma} -\,2\alpha \frac{k_{(\mu} P_{\nu)(\rho}k_{\sigma)}}{k^2}
           -\frac{2\alpha}{(\beta-2)^2} \frac{k_{\mu} k_{\nu}k_{\rho}k_{\sigma}}{k^4}
\Bigg].
\end{aligned}
\end{equation}
Here $A$ and $B$ are the form factors in the self-energy given in~\eqref{AB}. We derived the dressed propagator in $D$ dimensions in the general covariant gauge to facilitate its application to other research, while we see that the part proportional to $\alpha$ does not get dressed at all and therefore is irrelevant for our discussion about the non-trivial recovery of classical gravity at low energies. Moreover, it would be convenient if the propagator is transverse. We hence choose the Landau gauge~\cite{Galiza:1984ioc} (by setting $\alpha,\beta=0$) and let $D \to 4$ from now on so that the dressed propagator takes the concise form,
\begin{equation}\label{Landau_pro}
i\big[{}_{\mu\nu}\Delta_{\rho\sigma}^{(1)}\big]=\frac{2i}{k^{2}}\left[ \frac{1}{1+\frac{3\kappa^{2}}{k^{2}}A}\frac{P_{\mu\nu}P_{\rho\sigma}}{6}-\frac{1}{1-\frac{2\kappa^{2}}{k^{2}}B}Q_{\mu\nu\rho\sigma} \right].
\end{equation}
The first and the second terms in the bracket are the spin-0 and spin-2 components of the propagator as we mentioned in the introduction. We see that $k^{2}=0$ is a pole of the dressed propagator. In addition, it is a simple pole since $A/k^{2}$ and $B/k^{2}$ remain finite when $k^{2}\to 0$, as can be seen from~\eqref{A_limit} and~\eqref{B_limit}. Therefore, graviton stays massless at one loop.
However, additional poles do exist for non-perturbatively large momenta and we refer the reader to~\cite{Platania:2022gtt, Jimu:2024} for more details.
\section{The spin-2 and spin-0 residues match up}
\label{The spin-2 and spin-0 residues match up}
We now show the uniqueness of the low energy gravitational constant by investigating the dressed graviton propagator.

We first compute the limits regarding the function $L(k^{2})$. By its definition in~\eqref{L_def}, we have:
\begin{equation}\label{L_limit}
\lim_{k^{2} \to 0}L(k^{2})k^{2}=0, \qquad \quad\  \lim_{k^{2} \to 0}L(k^{2})=0, \qquad \quad\  \lim_{k^{2} \to 0}\frac{L(k^{2})}{k^{2}}=\int_{0}^{1}\mathrm{d}\alpha \frac{\alpha(1-\alpha)}{m^{2}}=\frac{1}{6m^{2}}.
\end{equation}
Using these results and the expressions of $A$, $B$ given in~\eqref{AB}, we can evaluate the limits:
\begin{equation}\label{A_limit}
\begin{aligned}
\lim_{k^{2} \to 0}\frac{A}{k^{2}}&=\lim_{k^{2} \to 0}\bigg\{-\frac{\Omega L(k^{2})}{4k^{2}}\frac{m^{4}}{9}+\left[ -\frac{\Omega m^{2}\Gamma_{E}}{6}\left( \xi-\frac{1}{6} \right)+\frac{\Omega m^{2}}{216}+\frac{c_{3\rm f}}{3} \right] \bigg\}\\
&=-\frac{\Omega m^{2}\Gamma_{E}}{6}\left( \xi-\frac{1}{6} \right)+\frac{c_{3\rm f}}{3},
\end{aligned}
\end{equation}
and
\begin{equation}\label{B_limit}
\begin{aligned}
\lim_{k^{2} \to 0}\frac{B}{k^{2}}&=\lim_{k^{2} \to 0} \bigg \{ -\frac{\Omega L(k^{2})}{480k^{2}}16m^{4}+\left[ \frac{\Omega m^{2}\Gamma_{E}}{4}\left( \xi-\frac{1}{6} \right)+\frac{\Omega m^{2}}{180}-\frac{c_{3\rm f}}{2} \right]  \bigg \}\\
&=\frac{\Omega m^{2}\Gamma_{E}}{4}\left( \xi-\frac{1}{6} \right)-\frac{c_{3\rm f}}{2}.
\end{aligned}
\end{equation}
Remarkably, this implies that the two seemingly unrelated tensorial components of the dressed propagator~\eqref{Landau_pro} have {\it identical} residues at the massless pole:
\begin{equation}\label{Z_def}
Z(\mu) \equiv \lim_{k^{2} \to 0}\frac{1}{1+\frac{3\kappa^{2}}{k^{2}}A}=\lim_{k^{2} \to 0}\frac{1}{1-\frac{2\kappa^{2}}{k^{2}}B}=\frac{1}{1+\kappa^{2}\left[ -\frac{\Omega m^{2}\Gamma_{E}}{2}\left( \xi-\frac{1}{6} \right)+c_{3\rm f} \right]},
\end{equation}
such that we have,
\begin{equation}\label{ratio}
\ \ \ \ \ \ \ \ i\big[{}_{\mu\nu}\Delta_{\rho\sigma}^{(1)}\big] \;\;\stackrel{k^{2} \to 0}{\longrightarrow}\;\; 
Z(\mu)\!\times\! i\big[{}_{\mu\nu}\Delta_{\rho\sigma}^{(0)}\big] 
\,.
\end{equation}
This shows that although there could be two different definitions for the running behavior of $G$ according to the two tensorial components, they run to the same value at low energies and therefore low energy gravitational constant is unique.
\section{Preparation: deriving the quantum-corrected gravitational potentials}
\label{Preparation: deriving the quantum-corrected gravitational potentials}
Now we turn to investigate the two gravitational potentials and verify that they are characterized by a unique gravitational constant at low energies. We calculate the one-loop corrections to the potentials induced by the massive scalar in this section and carry out the verification in the next section.

There are in general two approaches we can adopt to compute the quantum corrections to the potentials. One is using the {\it inverse scattering method} (see {\it e.g.}~\cite{Burns:2014bva, Bjerrum-Bohr:2002gqz, Bjerrum-Bohr:2014zsa, Bjerrum-Bohr:2016hpa, Bai:2016ivl, Chi:2019owc}), and the other is solving the {\it effective field equations} (see {\it e.g.}~\cite{Frob:2016xte, Park:2010pj, Marunovic:2011zw, Wang:2015eaa, Frob:2017smg, Miao:2024atw}). In general, the former approach is more suitable for obtaining gauge-invariant results while the latter can be more easily generalized to curved space~\cite{Miao:2017feh}. In this paper, we employ the latter approach. We will not come across gauge-dependence issues since only the loops of scalars are considered.
\subsection{Solving the effective field equations}
\label{Solving the effective field equation}
The effective action in which the massive scalar degree of freedom has been integrated out takes the form,
\begin{equation}
\Gamma_{h}=S_{h}-\frac{1}{2}\int \mathrm{d}^{4}x\int \mathrm{d}^{4}x' h^{\mu\nu}(x)\big[{}_{\mu\nu}\Sigma_{\rho\sigma}\big](x,x')h^{\rho\sigma}(x')\,.
\end{equation}
Here $S_{h}$ is the classical action of graviton given in~\eqref{expanded_action}. 
The consequent effective field equations in the presence of matter source take the form,
\begin{equation}\label{eff_eom}
-\mathcal{L}_{\mu \nu \rho \sigma}\kappa h^{\rho \sigma}(x)-\int d^{4}x' \big[{}_{\mu \nu}\Sigma_{\rho \sigma}\big](x,x')\kappa h^{\rho \sigma}(x')=\frac{\kappa^{2}}{2}T_{\mu \nu}(x)\,,
\end{equation}
in which $\mathcal{L}_{\mu\nu\rho\sigma}$ is given in~\eqref{lich}. For the source of the potentials, we consider a static point particle with mass $M$ at the origin. Thus,
\begin{equation}\label{source}
T_{\mu \nu}=M\delta^{3}(\vec{x})\delta_{\mu}^{0}\delta_{\nu}^{0}\,.
\end{equation}
We can then choose the Newtonian gauge in which the line element takes the form,
\begin{equation}\label{pp_def}
ds^2=-(1+2\Phi)dt^{2}+(1-2\Psi)(dx^{2}+dy^{2}+dz^{2})\,,
\end{equation}
where $\Phi$ and $\Psi$ are the two gravitational potentials.

We can solve~\eqref{eff_eom} with ease in momentum space. Contrary to the computation of the dressed propagator, here it is slightly more convenient to use $P_{\mu\nu\rho\sigma}$ instead of $Q_{\mu\nu\rho\sigma}$. We hence rewrite the self-energy~\eqref{finite_se} as,
\begin{equation}\label{OS_se_tilde}
-i\big[{}_{\mu\nu}\Sigma_{\rho\sigma}\big]=i\kappa^{2}\widetilde{A}P_{\mu\nu}P_{\rho\sigma}+i\kappa^{2}\widetilde{B}P_{\mu\nu\rho\sigma}\,,
\end{equation}
where
\begin{equation}\label{AB_tilde}
\widetilde{A}=A-\frac{B}{3}\,, \qquad \qquad \widetilde{B}=B\,.
\end{equation}
The Fourier transform of the effective field equations~\eqref{eff_eom} then take the form,
\begin{equation}\label{momentum_efe}
\frac{k^{2}}{2}(P_{\mu\nu}P_{\rho\sigma}-P_{\mu\nu\rho\sigma})h^{\rho\sigma}+\kappa^{2}(\widetilde{A}P_{\mu\nu}P_{\rho\sigma}+\widetilde{B}P_{\mu\nu\rho\sigma})h^{\rho\sigma}=\kappa \pi M\delta(k^{0})\delta^{0}_{\mu}\delta^{0}_{\nu}
\,.
\end{equation}
After some manipulations, the 00 component of~\eqref{momentum_efe} reads,
\begin{equation}\label{00}
-\|\vec{k}\|^2\Psi+\kappa^{2}(\widetilde{A}+\widetilde{B})\frac{\|\vec{k}\|^{4}}{k^{4}}\Phi+\kappa^{2}\left[ \widetilde{A}\left( \frac{\|\vec{k}\|^{2}}{k^{2}}-3 \right)+\widetilde{B}\frac{(k^{0})^{2}}{k^{2}} \right]\frac{\|\vec{k}\|^{2}}{k^{2}}\Psi=\frac{\kappa^2\pi M}{2}\delta(k^{0})
\,,
\end{equation}
and the 11+22+33 component reads,
\begin{equation}\label{11+22+33}
\begin{aligned}
-\|\vec{k}\|^2\Phi+(-3(k^{0})^2+\|\vec{k}\|^2)\Psi+\kappa^{2}&\left[ \widetilde{A}\left(\frac{\|\vec{k}\|^{2}}{k^{2}}-3 \right)+\widetilde{B}\frac{(k^{0})^{2}}{k^{2}} \right]\frac{\|\vec{k}\|^{2}}{k^{2}}\Phi\\
&+\kappa^{2}\left[ \widetilde{A}\left( \frac{\|\vec{k}\|^{2}}{k^{2}}-3 \right)^{2}+\widetilde{B}\left( \frac{(k^{0})^{4}}{k^{4}}+2 \right) \right]\Psi=0
\,.
\end{aligned}
\end{equation}
We can solve~\eqref{00} and~\eqref{11+22+33} perturbatively. Expand the potentials in the form,
\begin{equation}\label{expanded_h}
\Phi=\Phi^{(0)}+\kappa^{2}\Phi^{(1)} +{\cal O}(\kappa^4)
\,, \qquad \qquad \Psi=\Psi^{(0)}+\kappa^{2}\Psi^{(1)} +{\cal O}(\kappa^4)
\,,
\end{equation}
where the superscript (0) and (1) stand for the classical and quantum contributions respectively. At the leading order,~\eqref{00} and~\eqref{11+22+33} reduce to,
\begin{equation}
\begin{aligned}
-\|\vec{k}\|^2\Psi^{(0)}&=\frac{\kappa^{2} \pi M}{2}\delta(k^{0})\,,\\
-\|\vec{k}\|^2\Phi^{(0)}+(-3(k^{0})^2+\|\vec{k}\|^2)\Psi^{(0)}&=0
\,,
\end{aligned}
\end{equation}
which have the following solution,
\begin{equation}\label{leading_h}
\Phi^{(0)}=\Psi^{(0)}=-\frac{\kappa^{2}\pi M}{2\|\vec{k}\|^2}\delta(k^{0})\,.
\end{equation}
At the next-to-leading order,~\eqref{00} and~\eqref{11+22+33} take the form,
\begin{equation}\label{ntl_eom}
\begin{aligned}
-&\,\|\vec{k}\|^2\Psi^{(1)}+(\widetilde{A}+\widetilde{B})\frac{\|\vec{k}\|^{4}}{k^{4}}\Phi^{(0)}
+\left[ \widetilde{A}\left( \frac{\|\vec{k}\|^{2}}{k^{2}}-3 \right)+\widetilde{B}\frac{(k^{0})^{2}}{k^{2}} \right]\frac{\|\vec{k}\|^{2}}{k^{2}}\Psi^{(0)}=0\,,\\
\ \\
-&\,\|\vec{k}\|^2\Phi^{(1)}+(-3(k^{0})^2+\|\vec{k}\|^2)\Psi^{(1)}+\left[ \widetilde{A}\left(\frac{\|\vec{k}\|^{2}}{k^{2}}-3 \right)+\widetilde{B}\frac{(k^{0})^{2}}{k^{2}} \right]\frac{\|\vec{k}\|^{2}}{k^{2}}\Phi^{(0)}
 \\
&\ \hskip 5.8cm
+\bigg[ \widetilde{A}\left( \frac{\|\vec{k}\|^{2}}{k^{2}}-3 \right)^{2}+\widetilde{B}\left( \frac{(k^{0})^{4}}{k^{4}}+2 \right) \bigg]\Psi^{(0)}=0
\,.
\end{aligned}
\end{equation}
Thanks to the factor $\delta(k^{0})$ in $\Phi^{(0)}$ and $\Psi^{(0)}$, we can take $\|\vec{k}\|^{2}/k^{2} \to 1$ and $(k^{0})^{2}/k^{2} \to 0$ in~\eqref{ntl_eom}, after which it can be readily solved to give,
\begin{equation}\label{ntl_h}
\Phi^{(1)}=\frac{\widetilde{A}+3\widetilde{B}}{\|\vec{k}\|^{2}}\Phi^{(0)}\,, \qquad \qquad
\Psi^{(1)}=\frac{-\widetilde{A}+\widetilde{B}}{\|\vec{k}\|^{2}}\Psi^{(0)}
\,.
\end{equation}
Inserting~\eqref{leading_h} and~\eqref{ntl_h} into~\eqref{expanded_h} and reverting back to the form factors without tilde through~\eqref{AB_tilde}, we obtain the potentials,
\begin{equation}\label{momentum_pp}
\begin{aligned}
\Phi(x)&=\int \frac{\mathrm{d}^{4}k}{(2\pi)^{4}}e^{ik\cdot x}\left\{-\frac{\kappa^{2} \pi M\delta(k^{0})}{2\|\vec{k}\|^2}
\left[ 1+\frac{\kappa^{2}}{\|\vec{k}\|^{2}}\left( A+\frac{8}{3}B \right) \right]\right\}\,,\\
\Psi(x)&=\int \frac{\mathrm{d}^{4}k}{(2\pi)^{4}}e^{ik\cdot x}\left\{-\frac{\kappa^{2} \pi M\delta(k^{0})}{2\|\vec{k}\|^2}
\left[ 1-\frac{\kappa^{2}}{\|\vec{k}^{2}\|}\left( A-\frac{4}{3}B \right) \right]\right\}
\,.
\end{aligned}
\end{equation}
\section{Corrections to the two potentials converge at large distances}
\label{Corrections to the two potentials converge at large distances}
We now show that the quantum corrections to the two gravitational potentials converge as we go to large distances and therefore defining the low energy gravitational constant according to either potential does not affect the result.

Large distance in position space corresponds to $k^{2} \to 0$ in momentum space ($k$ is the momentum transfer between the two gravitating objects). From~\eqref{A_limit} and~\eqref{B_limit}, we have,
\begin{equation}\label{limA_limB}
\lim_{k^{2} \to 0}\frac{A}{k^{2}}=-\frac{2}{3}\lim_{k^{2} \to 0}\frac{B}{k^{2}}.
\end{equation}
Replacing the $\widetilde{A}$, $\widetilde{B}$ in~\eqref{ntl_h} with $A$, $B$ via~\eqref{AB_tilde} and using the above relation, the asymptotic values of the quantum corrections as $k^{2} \to 0$ are related to the tree-level potentials via,
\begin{equation}
\kappa^{2}\Phi^{(1)} \to \Phi^{(0)} \times 2\kappa^{2}\lim_{k^{2} \to 0}\frac{B}{k^{2}}\,, \qquad \qquad
\kappa^{2}\Psi^{(1)} \to \Psi^{(0)} \times 2\kappa^{2}\lim_{k^{2} \to 0}\frac{B}{k^{2}}\,.
\end{equation}
Since $\Phi^{(0)}=\Psi^{(0)}$, this shows the convergence of the quantum corrections at large distances and eliminates the ambiguity in defining the low energy gravitational constant. Specifically, by using the identity,
\begin{equation}\label{fourier_1}
\int \frac{\mathrm{d}^{3}k}{(2\pi)^{3}}e^{i\vec{k}\cdot \vec{x}} \frac{1}{\|\vec{k}\|^{2}}=\frac{1}{4\pi r},
\end{equation}
we can Fourier transform the tree-level part of~\eqref{momentum_pp} and get,
\begin{equation}
\Phi^{(0)}=\Psi^{(0)}=-\frac{GM}{r}.
\end{equation}
Recall that $\kappa^{2}=16\pi G$, the measured Newton's constant and the renormalized gravitational coupling $G$ as per the renormalization scheme are related by,

\begin{equation}
G_{N}=G\left(1+32\pi G\lim_{k^{2} \to 0}\frac{B}{k^{2}}\right).
\end{equation}
If we choose the on-shell renormalization scheme, as we will do in the next section, $\lim_{k^{2} \to 0}\frac{B}{k^{2}}=0$ and so $G_{N}$ and $G$ can be identified.

The above analysis shows that Bohr's correspondence principle holds nicely in gravity-massive matter systems. The interesting feature of this must-have property is that it relies on the connection between the seemingly unrelated spin-2 and spin-0 sectors, as can be seen from~\eqref{limA_limB}.
\section{Properties of the gravitational potentials in position space}
\label{Properties of the gravitational potentials in position space}
In the rest of the paper, we study the properties of the quantum-corrected gravitational potentials in more details. Specifically, we will express them analytically in terms of special functions, completing the previous study~\cite{Frob:2016xte} where only the asymptotic expressions are obtained due to the complexity of the integrals involved. This not only allows us to analyze the potentials across all distance scales, but also reveals a connection between pQG and QED through the structural similarity among their potentials. We will also show that our model can be constrained by experiments which results in a bound on the non-minimal coupling $\xi$.
\subsection{Setting up the on-shell scheme}
\label{Setting up the on-shell scheme}
Up to now we have not fixed the renormalization scheme yet in order to make our argument general. Next we shall adopt the the on-shell renormalization scheme which is convenient for comparison with experiments. In this scheme, we let $Z(\mu)$ defined in~\eqref{Z_def} equal to $1$ by choosing,
\begin{equation}\label{c_3f}
c_{3\rm f}=\frac{\Omega m^{2}\Gamma_{E}}{2}\left( \xi-\frac{1}{6} \right).
\end{equation}
In this way the renormalized gravitational coupling $G$ can be identified with the measured Newton's constant $G_{N}$, as we mentioned before.

In the on-shell scheme, the form factors $A$ and $B$ in the self-energy \eqref{finite_se} take particularly simple forms given by:
\begin{equation}\label{OS_AB}
\begin{aligned}
A&=-\frac{\Omega L(k^{2})}{4}\!\left[ k^{2}\left(\xi-\frac{1}{6}\right)\!+\!\frac{m^{2}}{3} \right]^{2}+\left[ -\frac{\Omega}{4}\left(\xi-\frac{1}{6}\right)^{2}\!+\!\frac{\Omega}{36}\left( \xi-\frac{1}{6} \right)\!+\!2c_{2\rm x} \right]k^{4}+\frac{\Omega m^{2}}{216}k^{2},\\
B&=-\frac{\Omega L(k^{2})}{480}(k^{2}+4m^{2})^{2}+\left( \frac{\Omega}{450}+c_{1\rm x} \right)k^{4}+\frac{\Omega m^{2}}{180}k^{2},
\end{aligned}
\end{equation}
where we have defined $c_{1\rm x}$ and $c_{2\rm x}$ as the finite counterterms through the relations:
\begin{equation}\label{cf_cx_relation}
c_{1\rm x}=c_{1\rm f}-\frac{\Omega\Gamma_{E}}{240}, \qquad \qquad \qquad \qquad c_{2\rm x}=c_{2\rm f}-\frac{\Omega\Gamma_{E}}{4}\left(\xi-\frac{1}{6}\right)^{2}.
\end{equation}
We note that when we compute the $k^{2} \to 0$ limit of $A/k^{2}$ and $B/k^{2}$ in~\eqref{A_limit} and~\eqref{B_limit}, the contributions from the $k^{2}$ terms in $A$ or $B$ cancel the leading order contributions from the non-local ({\it i.e.} $L(k^{2})$ dependent) terms. As a result, the self-energy becomes local at energies much below the scalar mass. As pointed out in \cite{Donoghue:2022chi}, this is a manifestation of the {\it decoupling} of the massive scalar in the infrared limit~\cite{Appelquist:1974tg}.
\subsection{Unified formulation for gravitational and Coulomb potentials}
\label{Unified formualtion for gravitational and Coulomb potentials}
We now perform the Fourier transform in~\eqref{momentum_pp} to obtain the potentials in position space. In this process we need the following identities:
\begin{align}
\int \frac{\mathrm{d}^{3}k}{(2\pi)^{3}}e^{i\vec{k}\cdot \vec{x}} L(\|\vec{k}\|^{2})&=-\frac{1}{2\pi r}\mathcal{J}_{0}(m,r), \label{fourier_L0}\\
\int \frac{\mathrm{d}^{3}k}{(2\pi)^{3}}e^{i\vec{k}\cdot \vec{x}} \frac{L(\|\vec{k}\|^{2})}{\|\vec{k}\|^{2}}&=\frac{1}{2\pi r}\mathcal{J}_{1}(m,r), \label{fourier_L1}\\
\int \frac{\mathrm{d}^{3}k}{(2\pi)^{3}}e^{i\vec{k}\cdot \vec{x}} \frac{L(\|\vec{k}\|^{2})}{(\|\vec{k}\|^{2})^{2}}&=-\frac{1}{2\pi r}\left( \mathcal{J}_{2}(m,r)-\frac{1}{12m^{2}} \right), \label{fourier_L2}
\end{align}
where $L(\|\vec{k}\|^{2})=\int_{0}^{1}\mathrm{d}\alpha\ln\left[1+\frac{\|\vec{k}\|^{2}}{m^{2}}\alpha(1-\alpha)\right]$ as we defined before and,
\begin{equation}
\mathcal{J}_{n}(m,r)=\left\{
\begin{aligned}
&(2m)^{2}\frac{K_{1}(z)}{z}\,, \qquad \qquad \qquad \qquad \qquad \ (n=0).\\
&(2m)^{2-2n}\Big[\mathrm{Ki}_{2n-2}(z)-\mathrm{Ki}_{2n}(z)\Big]\,, \qquad (n \ge 1).
\end{aligned}
\right.
\end{equation}
Here $K_{n}(z)$ is the Bessel function and $\mathrm{Ki}_{n}(z)$ is the Bickley function. We put the derivation of~\eqref{fourier_L0} --~\eqref{fourier_L2} in Appendix~\ref{Integrals}. There we used the techniques developed for the evaluation of the quantum-corrected Coulomb potential by~\cite{Frolov:2011rm}.

As we shall show below, by introducing the function $\mathcal{J}_{n}(m,r)$, we can write the gravitational and Coulomb potentials in a unified language. This will be enlightening for revealing the common structures shared by gravitational and electromagnetic forces.

Inserting the expressions of $A$ and $B$ given in~\eqref{OS_AB} into~\eqref{momentum_pp}, performing the Fourier transform via the integral identities above, and neglecting the delta function terms, we arrive at the quantum-corrected gravitational potentials sourced by a static point mass in Minkowski space, due to the one-loop effects of a massive scalar:
\begin{equation}\label{J_pp}
\begin{aligned}
\Phi (r)&=-\frac{GM}{r}\bigg\{ 1+\frac{G}{\pi}\left[ \left(\frac{1}{20}+\xi^{2}-\frac{\xi}{3}\right)\mathcal{J}_{0}-m^{2}\left(\frac{1}{15}+\frac{2\xi}{3}\right)\mathcal{J}_{1}+\frac{7m^{4}}{15}\mathcal{J}_{2} \right] \bigg\},\\
\Psi(r)&=-\frac{GM}{r}\bigg\{ 1-\frac{G}{\pi}\left[ \left(\frac{1}{60}+\xi^2-\frac{\xi}{3}\right)\mathcal{J}_{0}-m^{2}\left(\!-\frac{1}{5}+\frac{2\xi}{3}\right)\mathcal{J}_{1}-\frac{m^{4}}{15}\mathcal{J}_{2} \right] \bigg\}.
\end{aligned}
\end{equation}
For comparison, the quantum-corrected Coulomb potential which takes vacuum polarization into account reads~\cite{Uehling:1935uj, Frolov:2011rm},
\begin{equation}
\phi_C(r)=\frac{Q}{4\pi r}\left[ 1+\frac{e^{2}}{6\pi^{2}}( \mathcal{J}_{1}+2m_{e}^{2}\mathcal{J}_{2}) \right],
\end{equation}
where $Q$ is the electric charge of the source and $m_{e}$ here is the mass of electron. We thus see quite similar expressions for the two kinds of potentials. The function $\mathcal{J}_{0}$ which is absent in $\phi_C(r)$ somehow encodes the loop effects of a massless scalar on gravitation.

The similarity becomes more apparent at large distances. When $z\to \infty$, to sub-subleading order, we have the asymptotic forms,
\begin{equation}\label{long_K}
\begin{aligned}
K_{n}(z)&\sim \sqrt{\frac{\pi}{2z}}e^{-z}\left[ 1+\frac{4n^{2}-1}{8z}+\frac{(4n^{2}-1)(4n^{2}-9)}{128z^{2}} \right],\\
\mathrm{Ki}_{n}(z)&\sim \sqrt{\frac{\pi}{2z}}e^{-z}\left[ 1-\frac{4n+1}{8z}+\frac{3(16n^{2}+24n+3)}{128z^{2}} \right].
\end{aligned}
\end{equation}
The large-distance approximation of the gravitational potentials therefore reads,
\begin{equation}\label{long_pp}
\begin{aligned}
\Phi (r)|_{mr \gg 1}&=-\frac{GM}{r}\left\{1+G\sqrt{\frac{m}{\pi r^{3}}}e^{-2mr}\left[ \left(\xi-\frac{1}{4}\right)^{2}+\frac{3}{16mr}\left( \xi-\frac{1}{4} \right)\left( \xi+\frac{13}{12} \right) \right] \right\},\\
\Psi (r)|_{mr \gg 1}&=-\frac{GM}{r}\left\{1-G\sqrt{\frac{m}{\pi r^{3}}}e^{-2mr}\left[ \left(\xi-\frac{1}{4}\right)^{2}+\frac{3}{16mr}\left( \xi-\frac{1}{4} \right)\left( \xi+\frac{13}{12} \right) \right] \right\},
\end{aligned}
\end{equation}
which agrees with~\cite{Frob:2016xte}. Similarly, we have~\cite{Uehling:1935uj},
\begin{equation}
\phi_C(r)|_{m_{e}r \gg 1}=\frac{Q}{4\pi r}\left[ 1+\frac{e^{2}}{16}\frac{e^{-2m_{e}r}}{(\pi m_{e}r)^{\frac{3}{2}}} \right].
\end{equation}
We see that the quantum corrections to both kinds of potentials fall off in the same pattern $\sim\exp(-2mr)/r^{5/2}$ at large scales. Curiously, up to the order considered here, there is no quantum correction to the combination $\Phi(r)|_{mr \gg 1} + \Psi(r)|_{mr \gg 1}$, and the respective corrections to $\Phi(r)|_{mr \gg 1}$ and $\Psi(r)|_{mr \gg 1}$ vanish when $\xi=1/4$.

On the other hand, at short distances, the two kinds of potentials behave in quite different manners. When $z\to 0$, to subleading order, we have the asymptotic forms,
\begin{equation}\label{short_K}
\begin{aligned}
K_{0}(z)&\sim -\left[ \ln\left(\frac{z}{2}\right)+\gamma_{E} \right]-\frac{z^{2}}{4}\left[ \ln\left(\frac{z}{2}\right)+\gamma_{E}-1 \right], \qquad \qquad \mathrm{Ki}_{2}(z)\sim 1-\frac{\pi}{2}z,\\
K_{1}(z)&\sim \frac{1}{z}+\frac{z}{2}\left[\ln\left(\frac{z}{2}\right)+\gamma_{E}-\frac{1}{2}\right], \qquad \qquad \qquad \qquad \qquad \ \  \mathrm{Ki}_{4}(z)\sim \frac{2}{3}-\frac{\pi}{4}z.
\end{aligned}
\end{equation}
The short-distance approximation of the potentials then reads,
\begin{equation}\label{short_pp}
\begin{aligned}
\Phi (r)|_{mr \ll 1}=-\frac{GM}{r}\bigg\{ 1&+\frac{G}{\pi r^{2}}\left(\frac{1}{20}+\xi^{2}-\frac{\xi}{3}\right)\\
&+\frac{Gm^{2}}{\pi}\left[\left(\frac{1}{6}+2\xi^{2}\right)\Big(\ln(mr)+\gamma_{E}\Big)+\frac{1}{18}-\xi^{2}+\xi\right] \bigg\},\\
\Psi(r)|_{mr \ll 1}=-\frac{GM}{r}\bigg\{ 1&-\frac{G}{\pi r^{2}}\left(\frac{1}{60}+\xi^{2}-\frac{\xi}{3}\right)\\
&-\frac{Gm^{2}}{\pi}\left[\left(-\frac{1}{6}+2\xi^{2}\right)\Big(\ln(mr)+\gamma_{E}\Big)-\frac{2}{9}-\xi^{2}+\xi\right] \bigg\},
\end{aligned}
\end{equation}
which again agrees with~\cite{Frob:2016xte}. For comparison~\cite{Uehling:1935uj},
\begin{equation}
\phi_{C}(r)|_{m_{e}r \ll 1}=\frac{Q}{4\pi r}\left[ 1-\frac{e^{2}}{6}\left( \ln(m_{e}r)+\gamma_{E}+\frac{5}{6} \right) \right].
\end{equation}
We see that the Coulomb potential in this regime lacks the massless-scalar-induced inverse power terms in its gravitational counterparts.
\subsection{Experimental constraint}
\label{Experimental constraint}
From~\eqref{J_pp} we see the quantum corrections to $\Phi$ and $\Psi$ are different for most of the distance scales, which indicates a non-zero gravitational slip $\Sigma$ defined as $\Sigma \equiv \Phi-\Psi$. $\Sigma$ vanishes in GR and therefore serves as a probe for beyond-GR effects. Here we shall compare our theoretical prediction of $\Sigma$ with the experimental results.

According to~\eqref{long_pp}, the contributions to $\Sigma$ from heavy scalars 
(those satisfying $mr\gg 1$ on solar system scales) exponentially decay and hence have no significance at macroscopic scales. We thus focus on the effects of ultra-light and massless scalars, which have been proposed as dark matter~\cite{Antypas:2022asj,Friedrich:2019hev,Lee:2017qve,Boehm:2003hm} 
and dark energy~\cite{Glavan:2014uga,Glavan:2017jye,Belgacem:2021ieb,Belgacem:2022fui} candidates. Using~\eqref{short_pp}, we find the non-zero gravitational slip induced by such light scalars,
\begin{equation}
\frac{\Sigma}{\Phi_0} = \frac{2G}{\pi r^2}\left(\xi^2 - \frac{\xi}{3}+\frac{1}{30}\right)
   + \frac{2Gm^2}{\pi}\left[2\xi^2\big(\ln(mr)+\gamma_E\big) -\xi^2 + \xi -\frac{1}{12}\right] ,
\end{equation}
where $\Phi_0=\Psi_0=-\frac{GM}{r}$ are the tree-level potentials. 
For ultra-light fields considered here, the latter contribution is suppressed as $\sim (mr)^2\ln(mr)\ll 1$ and can thus be neglected.
On the other hand, solar system tests have shown that~\cite{Bertotti:2003rm},
\begin{equation}
\frac{\Sigma}{\Phi_0} < 2\times 10^{-5}.
\end{equation}
Combining the above two results, we obtain a constraint on the non-minimal coupling $\xi$,
\begin{equation}
|\xi|  < \sqrt{10\pi}\frac{r}{l_P}\!\times\!10^{-3} \sim 5\times 10^{43},
\label{bound on xi}
\end{equation}
where $l_P = \sqrt{G} \approx 1.6\times 10^{-35}\,{\rm m}$ is the Planck length and we took $r\approx 1.5\times 10^{11}\,{\rm m}$ which is the Sun-Earth distance. The constraint can be strengthened if the field is in a highly excited state, or 
 if there is a large number of scalars running in the loop ($N\gg 1$). 
In the latter case, $|\xi| \lesssim 5\times 10^{43}/\sqrt{N}$. In any case, the constraint is very weak;
nevertheless, it is worth mentioning as it is a constraint established by perturbative quantum gravity through the investigation of the one-loop vacuum fluctuations of scalar matter in Minkowski space.
To our knowledge, no such constraints have been claimed by researching loop effects in quantum gravity before.
We also note that it would be of interest to revisit the upper bound~(\ref{bound on xi}) for 
scalar fields in highly excited states, such as those generated during the primordial
 inflation~\cite{Glavan:2014uga,Glavan:2017jye,Belgacem:2021ieb,Belgacem:2022fui,Adams:2022pbo}.
\section{Discussion}
\label{Discussion}
In this work, we considered the potential ambiguity in defining the gravitational constant at low energies when gravity is coupled to massive quantum fields. The issue arises from the fact that there is not a unique definition of the running behavior of the gravitational constant which can account for different kinds of interactions in perturbative quantum gravity (pQG). Though this reflects the limitation of pQG in probing the high energy regime of quantum gravity, we showed that it is reliable in bridging the high energy regime with the low energy one since the different definitions of the running behavior converge at low energies. 
This result can be viewed as a generalization of the Appelquist-Carazzone 
decoupling theorem~\cite{Appelquist:1974tg} -- originally shown to hold for matter fields  -- 
to perturbative quantum gravity.

We focused on the one-loop effects of a massive scalar and showed that the above is particularly true for the two running couplings defined according to the spin-2 and spin-0 components of the dressed graviton propagator, as the two components have equal residues at the massless pole. Considering the orthogonality between the spin-2 and spin-0 tensor structures, such a relation is somewhat surprising. It would be interesting to show whether this equivalence between their residues continues to hold at higher loop orders and for massive fields with spin.

In particular, we would like to know whether it is a corollary of the symmetries such as general covariance. If so, it would mean that the form of the scattering amplitudes in pQG are more severely constrained by symmetry than previously thought.

Since we probe gravity through quantities like the gravitational potentials (we probe $\Phi$ and $\Phi+\Psi$ by observing the motion of non-relativistic and relativistic particles respectively), we calculated the quantum corrections to them induced by the massive scalar and showed that the above connection between the spin-2 and spin-0 sectors allows us to operationally define the low energy gravitational constant according to the asymptotic behavior of either potential as the result is the same.

We also found that the gravitational potentials in our case as well as the quantum-corrected Coulomb potential in QED can be expressed concisely in terms of the special function $\mathcal{J}_{n}(m,r)$ we defined in the main text. Moreover, the corrections to both kinds of potentials decay as $\sim\exp(-2mr)/r^{5/2}$ in large distances. This shows that Newton's law and Coulomb's law are not only similar at the classical level, but also at the quantum level. It would be interesting to find out the exact reason behind this similarity and see whether it shows up in other force laws.

Since the quantum effects of the scalar lead to a non-zero gravitational slip, we compared it with the experiments conducted on the Solar System scale and obtained a constraint on the non-minimal coupling parameter: $|\xi| \lesssim 5\times 10^{43}/\sqrt{N}$, where $N$ is the number of ultra-light or massless scalars. Though it is a weak bound due to the nature of perturbative calculation, it is established on the relatively reliable ground of pQG and may be useful in reducing the parameter space of certain conjectures.
\acknowledgments
The authors thank Riley Kavanagh whose master's thesis~\cite{Kavanagh:2023} was
used for various aspects of this work.
This work is part of the Delta ITP consortium,
a program of the Netherlands Organisation for Scientific Research (NWO) that is
funded by the Dutch Ministry of Education, Culture and Science (OCW) — NWO
project number 24.001.027.
\appendix
\section{Feynman rules and amplitudes}
\label{Feynman rules and amplitudes}
Here we give the intermediate steps in the derivation of the self-energy~\eqref{finite_se}. From~\eqref{expanded_action} we can read off the Feynman rules for the three-point and four-point vertices,
\begin{figure}[H]
\centering
\subfigure{
\begin{tikzpicture}[baseline=-2pt]
\begin{feynman}
\vertex (a);
\vertex [left=of a] (b);
\vertex [above right=of a] (d);
\vertex [below right=of a] (c); 
\diagram{
(b) -- [gluon, momentum'={[arrow shorten=0.25]$k$}, edge label=$\mu\nu$] (a);
(c) -- [momentum={[arrow shorten=0.25]$p_{1}$}] (a) -- [momentum={[arrow shorten=0.25]$p_{2}$}] (d);
};
\end{feynman}
\end{tikzpicture}
}
\begin{equation}\label{Appendix A: cubic vertex}
=-i\kappa\left \{ p_{1}^{(\mu}p_{2}^{\nu)}-\frac{\eta^{\mu\nu}}{2}(p_{1}\cdot p_{2}+m^{2})-\xi \left[ \eta^{\mu\nu}(p_{1}-p_{2})^{2}-(p_{1}-p_{2})^{\mu}(p_{1}-p_{2})^{\nu} \right] \right \}\,. \hspace{0.5cm}
\end{equation}
\ \\
\ \\
\ \\
\subfigure{
\begin{tikzpicture}[baseline=-2pt]
\begin{feynman}
\vertex (a);
\vertex [left=of a] (b);
\vertex [right=of a] (c);
\vertex [above left=of a] (d);
\vertex [above right=of a] (e);
\diagram{
(b) -- [gluon, momentum'={[arrow shorten=0.25]$k$}, edge label=$\mu\nu$] (a) -- [gluon, momentum'={[arrow shorten=0.25]$k$}, edge label=$\rho\sigma$] (c);
(d) -- [momentum={[arrow shorten=0.25, arrow distance=2mm, label distance=-1pt]$p$}] (a) -- [momentum={[arrow shorten=0.25, arrow distance=2mm, label distance=-1pt]$p$}] (e);
};
\end{feynman}
\end{tikzpicture}
}
\begin{equation}\label{Appendix A: quartic vertex}
=\frac{i\kappa^{2}}{2}\left[ p^{\mu}p^{\nu}\eta^{\rho\sigma}+p^{\rho}p^{\sigma}\eta^{\mu\nu}-\left(\frac{1}{2}\eta^{\mu\nu}\eta^{\rho\sigma}+\eta^{\mu(\rho}\eta^{\sigma)\nu}\right)(p^{2}+m^{2})-\xi k^{2}(P^{\mu\nu}P^{\rho\sigma}-P^{\mu\nu\rho\sigma}) \right]
\,.
\end{equation}
\end{figure}

\noindent The momentum specification in the four-point vertex is not the most general one, but it suffices for our calculation at one-loop order. The amplitudes of diagrams (a) and (b) in Figure~\ref{figure_se} are then given by,
\begin{equation}
\begin{aligned}
\mathcal{M}_{\mu\nu,\rho\sigma}^{(a)}&=\frac{1}{2}\int\frac{\mathrm{d}^{D}p}{(2\pi)^{D}}(-i\kappa)\left \{ p_{(\mu}(p+k)_{\nu)}-\frac{\eta_{\mu\nu}}{2}[p\cdot (p+k)+m^{2}]-\xi k^{2}P_{\mu\nu} \right\}\frac{-i}{p^{2}+m^{2}}\\
&\qquad \qquad \times\frac{-i}{(p+k)^{2}+m^{2}}(-i\kappa)\left \{ p_{(\rho}(p+k)_{\sigma)}-\frac{\eta_{\rho\sigma}}{2}[p\cdot (p+k)+m^{2}]-\xi k^{2}P_{\rho\sigma} \right\},
\end{aligned}
\end{equation}
and
\begin{equation}
\begin{aligned}
\mathcal{M}_{\mu\nu,\rho\sigma}^{(b)}=\frac{1}{2}\int\frac{\mathrm{d}^{D}p}{(2\pi)^{D}}\frac{i\kappa^{2}}{2}\bigg[ p_{\mu}&p_{\nu}\eta_{\rho\sigma}+p_{\rho}p_{\sigma}\eta_{\mu\nu}-\left(\frac{1}{2}\eta_{\mu\nu}\eta_{\rho\sigma}+\eta_{\mu(\rho}\eta_{\sigma)\nu}\right)(p^{2}+m^{2})\\
&-\xi k^{2}(P_{\mu\nu}P_{\rho\sigma}-P_{\mu\nu\rho\sigma}) \bigg]\frac{-i}{p^{2}+m^{2}}.
\end{aligned}
\end{equation}
After some strenuous manipulations, we can put them into the form,
\begin{equation}
\begin{aligned}
\mathcal{M}_{\mu\nu,\rho\sigma}^{(a)}&=\kappa^{2}\left[\frac{4(D-2)m^{2}-(D^{2}-2D-2)k^2}{16(D^{2}-1)}P_{\mu\nu}P_{\rho\sigma}+\frac{4(D-2)m^{2}-k^{2}}{8(D^{2}-1)}P_{\mu\nu\rho\sigma}\right]I_{1}\\
&\quad \ +\kappa^{2}\left[\frac{(Dk^{2}-4m^{2})^{2}-2(D+1)k^4}{32(D^{2}-1)}P_{\mu\nu}P_{\rho\sigma}+\frac{(k^{2}+4m^{2})^{2}}{16(D^2-1)}P_{\mu\nu\rho\sigma}\right]I_{2}\\
&\quad \ +\frac{\kappa^{2}\xi (D-2)k^{2}}{2(D-1)}P_{\mu\nu}P_{\rho\sigma}I_{1}+\kappa^{2}\xi\left[ \frac{\xi k^{4}}{2}-\frac{(D-2)k^{4}}{4(D-1)}+\frac{m^{2}k^{2}}{D-1}  \right]P_{\mu\nu}P_{\rho\sigma}I_{2}\\
&\quad \ +\frac{\kappa^{2}m^{2}}{4D}(\eta_{\mu\nu}\eta_{\rho\sigma}-2\eta_{\mu(\rho}\eta_{\sigma)\nu})I_{1},\\
\end{aligned}
\label{Appendix A: self energy a}
\end{equation}
and
\begin{equation}
\mathcal{M}_{\mu\nu,\rho\sigma}^{(b)}=-\frac{\kappa^{2}m^{2}}{2D}\eta_{\mu\nu}\eta_{\rho\sigma}I_{1}-\frac{\kappa^{2}\xi k^{2}}{4}(P_{\mu\nu}P_{\rho\sigma}-P_{\mu\nu\rho\sigma})I_{1},
\label{Appendix A: self energy b}
\end{equation}
where the projectors $P_{\mu\nu}$ and $P_{\mu\nu\rho\sigma}$ are defined in~\eqref{projector_def}, and the integrals $I_{1}$ and $I_{2}$ defined in~\eqref{I1I2_def} can be evaluated,
\begin{equation}\label{Appendix: I1I2_def}
\begin{aligned}
I_{1}& =\! \int \!\frac{\mathrm{d}^{D}p}{(2\pi)^{D}} \frac{1}{p^{2}+m^{2}}=i\Omega m^{2}\mu^{D-4} \left ( \frac{1}{D\!-\!4}+\Gamma_{E} \right )+\mathcal{O}(D\!-\!4)\,,\\
I_{2}& =\! \int \!\frac{\mathrm{d}^{D}p}{(2\pi)^{D}} \frac{1}{(p^{2}\!+\!m^{2})[(p\!+\!k)^{2}\!+\!m^{2}]}=-i\Omega\mu^{D-4} \left ( \frac{1}{D\!-\!4}+\Gamma_{E}+\frac{L(k^{2})+1}{2} \right )+\mathcal{O}(D\!-\!4)
\,.
\end{aligned}
\end{equation}
The shorthand notations $\Omega$, $\Gamma_{E}$ and $L(k^{2})$ are defined in~\eqref{shorthand_def} and \eqref{L_def}. By expanding the counterterms in~\eqref{counterterms} to quadratic order in $h_{\mu\nu}$, we can find their contributions to diagram (c) in Figure~\ref{figure_se},
\begin{equation}\label{counter_amplitudes}
\begin{aligned}
\mathcal{M}_{\mu\nu,\rho\sigma}^{(C^{2})}&=\frac{2ic_{1}\kappa^{2}(D-3)}{D-2}k^{4}\left( P_{\mu\nu\rho\sigma}-\frac{1}{D-1}P_{\mu\nu}P_{\rho\sigma} \right),\\
\mathcal{M}_{\mu\nu,\rho\sigma}^{(R^{2})}&=2ic_{2}\kappa^{2}k^{4}P_{\mu\nu}P_{\rho\sigma},\\
\mathcal{M}_{\mu\nu,\rho\sigma}^{(R)}&=\frac{ic_{3}\kappa^{2}}{2}k^{2}\left( P_{\mu\nu}P_{\rho\sigma}-P_{\mu\nu\rho\sigma} \right),\\
\mathcal{M}_{\mu\nu,\rho\sigma}^{(\Lambda)}&=\frac{i\Lambda\kappa^{2}}{4}(\eta_{\mu\nu}\eta_{\rho\sigma}+2\eta_{\mu(\rho}\eta_{\sigma)\nu}).
\end{aligned}
\end{equation}
We see that, unlike other counterterm amplitudes, $\mathcal{M}_{\mu\nu,\rho\sigma}^{(\Lambda)}$ is not transverse, and indeed it cancels the non-transverse part of $\mathcal{M}_{\mu\nu,\rho\sigma}^{(a)}+\mathcal{M}_{\mu\nu,\rho\sigma}^{(b)}$ exactly. Combining the above amplitudes, one obtains the primitive self-energy~\eqref{self energy: primitive}, which after adding the counterterms~\eqref{counterterms},~\eqref{c_123}
and taking the $D \to 4$ limit, yields the renormalized self-energy~\eqref{finite_se}.
\section{Integrals}
\label{Integrals}
In this appendix we evaluate the integrals in~\eqref{fourier_L0},~\eqref{fourier_L1} and~\eqref{fourier_L2} using contour integration.
\begin{equation}\label{contour_integral}
\begin{aligned}
\int \frac{\mathrm{d}^{3}k}{(2\pi)^{3}}&e^{i\vec{k}\cdot \vec{x}} \frac{L\big(\|\vec{k}\|^{2}\big)}{(\|\vec{k}\|^{2})^{n}}
=\int \frac{\mathrm{d}^{3}k}{(2\pi)^{3}} \frac{e^{i\vec{k}\cdot \vec{x}}}{(\|\vec{k}\|^{2})^{n}}
\int_{0}^{1}\mathrm{d}\alpha\ln\left[1+\frac{\|\vec{k}\|^{2}}{m^{2}}\alpha(1-\alpha)\right]\\
&=\frac{1}{(2\pi)^{2}}\int_{0}^{1}\mathrm{d}\alpha\int_{0}^{\infty}\mathrm{d}\|\vec{k}\|\int_{-1}^{1}\mathrm{d}(\cos\theta)\,e^{i\|\vec{k}\|r
\cos\theta}\|\vec{k}\|^{2-2n}\ln\left[1+\frac{\|\vec{k}\|^{2}}{m^{2}}\alpha(1-\alpha)\right]\\
&=\frac{1}{(2\pi)^{2}ir}\int_{0}^{1}\mathrm{d}\alpha\int_{0}^{\infty}\mathrm{d}\|\vec{k}\|\,(e^{i\|\vec{k}\|r}-e^{-i\|\vec{k}\|r})\|\vec{k}\|^{1-2n}
\ln\left[1+\frac{\|\vec{k}\|^{2}}{m^{2}}\alpha(1-\alpha)\right]\\
&=\frac{1}{(2\pi)^{2}ir}\int_{0}^{1}\mathrm{d}\alpha\int_{-\infty}^{\infty}\mathrm{d}\|\vec{k}\|\,e^{i\|\vec{k}\|r}\|\vec{k}\|^{1-2n}
\ln\left[1+\frac{\|\vec{k}\|^{2}}{m^{2}}\alpha(1-\alpha)\right].
\end{aligned}
\end{equation}
Now we take the trickiest $n=2$ case as an example to illustrate the idea. The integrand possesses two branch cuts along the imaginary line due to the factor $\ln\left[1+\frac{\|\vec{k}\|^{2}}{m^{2}}\alpha(1-\alpha)\right]$. It also has a pole at the origin due to the factor $\|\vec{k}\|^{-3}$. We thus consider the contour composed of $\gamma_{1} \to \gamma_{2} \to \cdots \to \gamma_{8}$ shown in Figure~\ref{figure_contour}.
\begin{figure}[H]
\centering
\begin{tikzpicture}
\begin{feynman}
\vertex (a) at (-5, 0);
\vertex (b) at (-4,0);
\vertex (bc) at (-1.6,0);
\vertex (c) at (0,0) {\Large $\times$};
\vertex (d) at (4, 0);
\vertex (e) at (5,0);
\vertex (f) at (4*0.087156,4*0.996195);
\vertex (j) at (-4*0.087156,4*0.996195);
\vertex (g) at (4*0.087156, 1.5);
\vertex (i) at (-4*0.087156, 1.5);
\vertex (h) at (0,1.5-4*0.087156);
\vertex (h_2) at (0,1.5-4*0.087156+0.02) {$\blacktriangleleft$};
\vertex (k) at (0,4.54);
\vertex (l) [dot] at (0,1.5-4*0.087156+0.2) {\ };
\vertex (m) [dot] at (0,-1.5+4*0.087156-0.2) {\ };
\vertex (n) at (0,-2.2);
\vertex (sc1) at (-4*0.087156,0);
\vertex (sc2) at (0,4*0.087156);
\vertex (sc2_2) at (0.05,4*0.087156) {$\blacktriangleright$};
\vertex (sc3) at (4*0.087156,0);
\vertex (x) at (4.89,0) {\large $>$};
\vertex (v) at (0,4.455) {\large \rotatebox{90}{$>$}};
\vertex (w) at (4.85,-0.5) {$\mathrm{Re}(|\vec{k}|)$};
\vertex (u) at (0.8,4.35) {$\mathrm{Im}(|\vec{k}|)$};
\vertex (z) at (1.2,1.5-4*0.087156+0.2) {$\frac{im}{\sqrt{\alpha(1-\alpha)}}$};
\vertex (y) at (1.2,-1.5+4*0.087156-0.2) {$\frac{-im}{\sqrt{\alpha(1-\alpha)}}$};
\vertex (g1) at (-2.2,-0.5) {\large $\gamma_{1}$};
\vertex (g2) at (0.42,0.42) {\large $\gamma_{2}$};
\vertex (g3) at (2.2,-0.5) {\large $\gamma_{3}$};
\vertex (g4) at (3.4,3) {\large $\gamma_{4}$};
\vertex (g5) at (0.9,2.7) {\large $\gamma_{5}$};
\vertex (g6) at (-0.28,0.95) {\large $\gamma_{6}$};
\vertex (g7) at (-0.9,2.7) {\large $\gamma_{7}$};
\vertex (g8) at (-3.4,3) {\large $\gamma_{8}$};
\diagram {
(a) -- (b) -- [thick, fermion] (sc1) -- [thick, quarter left] (sc2) -- [thick, quarter left] (sc3) -- [thick, fermion] (d) -- (e);
(sc1) -- [thin] (sc3);
(d) -- [thick, fermion, out=90, in=-5] (f);
(j) -- [thick, fermion, out=-175, in=90] (b);
(f) -- [thick, fermion] (g) -- [thick, quarter left] (h) -- [thick, quarter left] (i) -- [thick, fermion] (j);
(k) -- [thin, boson] (l) -- [thin] (m) -- [thin, boson] (n);
};
\end{feynman}
\end{tikzpicture}
\caption{The contour used to evaluate the $n=2$ case of~\eqref{contour_integral}. It comprises eight pieces and bypasses both the branch cut along the imaginary line and the pole at the origin. $\gamma_{2}$ and $\gamma_{6}$ are infinitesimal semicircles. When $n=0$ or $n=1$, there's no pole at the origin so we do not need to curve the contour like $\gamma_{2}$. \label{figure_contour}}
\end{figure}

\noindent There is no pole enclosed in the whole contour, so the integral over it gives zero. This means~\eqref{contour_integral}, represented schematically as $\int_{\gamma_{1}+\gamma_{3}}$, is given by $-(\int_{\gamma_{4}+\gamma_{8}}+\int_{\gamma_{5}+\gamma_{7}}+\int_{\gamma_{6}}+\int_{\gamma_{2}})$. The contribution from the two arcs $\int_{\gamma_{4}+\gamma_{8}}$ vanishes due to the factor $e^{i\|\vec{k}\|r}$ in the integrand. $\int_{\gamma_{6}}$ also vanishes since it scales as $\sim \rho \ln(\rho)$ where $\rho$ is the radius of the semicircle. However, despite the size of $\gamma_{2}$ being infinitesimally small, $\int_{\gamma_{2}}$ does not vanish, as the following computation shows:
\begin{equation}
\begin{aligned}
\int_{\gamma_{2}}\bigg|_{n=2}&=\frac{1}{(2\pi)^{2}ir}\int_{0}^{1}\mathrm{d}\alpha\int_{\gamma_{2}}\mathrm{d}\|\vec{k}\|\,e^{i\|\vec{k}\|r}
\frac{1}{\|\vec{k}\|^{3}}\ln\left[1+\frac{\|\vec{k}\|^{2}}{m^{2}}\alpha(1-\alpha)\right]\\
&=\frac{1}{(2\pi)^{2}ir}\int_{0}^{1}\mathrm{d}\alpha \lim_{\rho \to 0} \int_{\pi}^{0}\mathrm{d}\theta\,(i\rho e^{i\theta})e^{i\rho e^{i\theta}r}\frac{1}{(\rho e^{i\theta})^{3}}\ln\left[1+\frac{(\rho e^{i\theta})^{2}}{m^{2}}\alpha(1-\alpha)\right]\\
&=\frac{1}{(2\pi)^{2}r}\int_{0}^{1}\mathrm{d}\alpha \lim_{\rho \to 0} \int_{\pi}^{0}\mathrm{d}\theta\,\frac{1}{\rho^{2} e^{2i\theta}}\frac{(\rho e^{i\theta})^{2}}{m^{2}}\alpha(1-\alpha)\\
&=\frac{1}{(2\pi)^{2}r}\int_{0}^{1}\mathrm{d}\alpha\,\frac{(-\pi)}{m^{2}}\alpha(1-\alpha)\\
&=-\frac{1}{24\pi m^{2}r}.
\end{aligned}
\end{equation}
Regarding the effect on the potentials, this contribution, stemming from the non-local part of the self-energy, will cancel those from the $k^{2}$ terms, which is a position space manifestation of the decoupling of the scalar we mentioned in the main text. Using $\ln(-|x| \pm i\epsilon)=\ln(|x|) \pm i\pi$ for $\epsilon \to 0^{+}$, we can evaluate $\int_{\gamma_{5}+\gamma_{7}}$ for a general $n$,
\begin{equation}
\begin{aligned}
\int_{\gamma_{5}+\gamma_{7}}&=\frac{1}{(2\pi)^{2}ir}\int_{0}^{1}\mathrm{d}\alpha\int_{\gamma_{5}+\gamma_{7}}\mathrm{d}\|\vec{k}\|
\,e^{i\|\vec{k}\|r}\|\vec{k}\|^{1-2n}\ln\left[1+\frac{\|\vec{k}\|^{2}}{m^{2}}\alpha(1-\alpha)\right]\\
&=\frac{1}{(2\pi)^{2}ir}\int_{0}^{1}\mathrm{d}\alpha\int_{\frac{im}{\sqrt{\alpha(1-\alpha)}}}^{i\infty}\mathrm{d}\|\vec{k}\|
\, e^{i\|\vec{k}\|r}\|\vec{k}\|^{1-2n}(-2i\pi)\\
&=\frac{(-1)^{n}}{2\pi r}\int_{0}^{1}\mathrm{d}\alpha\int_{\frac{m}{\sqrt{\alpha(1-\alpha)}}}^{\infty}\mathrm{d}t\, e^{-tr}t^{1-2n}\\
&=\frac{(-1)^{n}}{2\pi r}\int_{2m}^{\infty}\mathrm{d}t\,e^{-tr}t^{1-2n}\int_{\frac{1}{2}-\frac{1}{2}\sqrt{1-\frac{4m^{2}}{t^{2}}}}^{\frac{1}{2}+\frac{1}{2}\sqrt{1-\frac{4m^{2}}{t^{2}}}}\mathrm{d}\alpha\\
&=\frac{(-1)^{n}}{2\pi r}\int_{2m}^{\infty}\mathrm{d}t\,\frac{\sqrt{t^{2}-4m^2}}{t^{2n}}e^{-rt}\\
&=\frac{(-1)^{n}}{2\pi r}\mathcal{J}_{n}(m,r)\,,
\end{aligned}
\end{equation}
where $\mathcal{J}_{n}(m,r)$ is defined as,
\begin{equation}\label{Jn_def}
\mathcal{J}_{n}(m,r) \equiv \int_{2m}^{\infty}\mathrm{d}t\,\frac{\sqrt{t^{2}-4m^2}}{t^{2n}}e^{-rt}.
\end{equation}

Next we evaluate the integral in~\eqref{Jn_def} following the approach of~\cite{Frolov:2011rm} which evaluated similar integrals when trying to obtain the analytic formula for the quantum-corrected Coulomb potential. Define $z \equiv 2mr$. After a change of variable $t=2m\cosh(x)$,~\eqref{Jn_def} can be recast into the form,
\begin{equation}\label{Jn_rewritten}
\mathcal{J}_{n}(m,r)=(2m)^{2-2n}\int_{0}^{\infty}\mathrm{d}x\,\frac{\cosh^{2}(x)-1}{\cosh^{2n}(x)}e^{-z\cosh(x)}.
\end{equation}
For $\mathcal{J}_{0}(m,r)$, using the following properties of the modified Bessel function of the second kind,
\begin{equation}
\begin{aligned}
K_{0}(z)&=\int_{0}^{\infty}\mathrm{d}x\,e^{-z\cosh(x)},\\
K_{0}''(z)&=-\frac{K_{1}(z)}{z}+K_{2}(z),\\
K_{1}(z)&=-\frac{z}{2}\Big[K_{0}(z)-K_{2}(z)\Big],
\end{aligned}
\end{equation}
we get,
\begin{equation}\label{analytic_J0}
\mathcal{J}_{0}(m,r)=(2m)^{2}\Big[K_{0}''(z)-K_{0}(z)\Big]=(2m)^{2}\frac{K_{1}(z)}{z}.
\end{equation}
To evaluate $\mathcal{J}_{n}(m,r)$ for $n \ge 1$, we introduce the Bickley function $\mathrm{Ki}_{n}(z)$ which can be defined recursively as the integral of Bessel function,
\begin{equation}
\mathrm{Ki}_{0}(z) \equiv K_{0}(z), \quad \mathrm{and} \quad
\mathrm{Ki}_{n}(z) \equiv \int_{z}^{\infty} \mathrm{d}z'\,\mathrm{Ki}_{n-1}(z') \quad \mathrm{for} \quad n\ge1.
\end{equation}
$\mathrm{Ki}_{n}(z)$ has integral representation,
\begin{equation}
\mathrm{Ki}_{n}(z)=\int_{0}^{\infty}\mathrm{d}x\,\frac{e^{-z\cosh(x)}}{\cosh^{n}(x)}.
\end{equation}
Comparing this with~\eqref{Jn_rewritten} gives,
\begin{equation}\label{analytic_Jn}
\begin{aligned}
\mathcal{J}_{n}(m,r)&=(2m)^{2-2n}\Big[\mathrm{Ki}_{2n-2}(z)-\mathrm{Ki}_{2n}(z)\Big]
,
\qquad (n \ge 1).
\end{aligned}
\end{equation}
Combining the above results, we reach the identities~\eqref{fourier_L0}, \eqref{fourier_L1} and~\eqref{fourier_L2}.

\end{document}